\documentclass[preprint,3p, twocolumn]{elsarticle}
\usepackage[fleqn]{amsmath}
\usepackage{amssymb,amsthm}
\usepackage{mathtools}
\usepackage{mathrsfs}
\usepackage{bm}
\usepackage{slashed}	
\usepackage{cancel}
\usepackage{soul}
\usepackage{color}
\usepackage{xcolor}
\usepackage[pdftex,hypertexnames=false,linktocpage=true]{hyperref}
\hypersetup{colorlinks=true,linkcolor=red,anchorcolor=darkgreen,citecolor=green!50!blue,filecolor=darkgreen,urlcolor=green,
bookmarksnumbered=true,
pdfview=FitB
}
\usepackage{lineno}
\biboptions{sort&compress}
\usepackage{verbatim}
\usepackage{fullpage}
\usepackage{graphicx}
\usepackage{subfig} 	
\usepackage{relsize}	
\usepackage{float}
\usepackage{array}
\usepackage{booktabs}
\usepackage{arydshln}
\usepackage{multirow}

\def\arraystretch{1.3}

\makeatletter

\newcommand{\Rmnum}[1]{\expandafter\@slowromancap\romannumeral #1@}
\makeatother

\colorlet{darkgreen}{green!60!black}
\colorlet{brightyellow}{yellow!75!red}
\colorlet{orange}{red!50!yellow}
\colorlet{darkblue}{blue!60!black}
\colorlet{darkred}{red!80!black}
\colorlet{greenblue}{green!50!blue}

\begin{document}
\begin{frontmatter}

\title{Extending light-front holographic QCD using the 't Hooft Equation}

\author[mau]{Mohammad Ahmady}
%
\author[nitj]{Harleen Dahiya}
%
\author[nitj]{Satvir Kaur}
%
\author[imp,cas]{Chandan Mondal\corref{c1}}
%
\author[au]{Ruben Sandapen}  
%
\author[ptu]{Neetika Sharma}  
%
\cortext[c1]{Corresponding author}
\address[mau]{Department of Physics, Mount Allison University,  Sackville, New Brunswick,  E4L 1E6, Canada.}
\address[nitj]{Department of Physics, Dr. B. R. Ambedkar National Institute of Technology, Jalandhar 144011, India}
\address[imp]{Institute for Modern Physics, Chinese Academy of Sciences, Lanzhou-730000, China} 
\address[cas]{School of Nuclear Science and Technology, University of Chinese Academy of Sciences, Beijing 100049, China}
\address[au]{Department of Physics, Acadia University, Wolfville, Nova Scotia, B4P 2R6, Canada.}
\address[ptu]{ Department of Physical Sciences, 
	I K Gujral Punjab Technical University,
	Kapurthala-144603, Punjab, India.}

\begin{abstract}

\noindent We show the 't Hooft Equation and the light-front holographic Schr\"odinger Equation are complementary to each other in governing the transverse and longitudinal dynamics of colour confinement in quark-antiquark mesons. Together, they predict remarkably well the light, heavy-light and heavy-heavy meson spectroscopic data. The universal emerging hadronic scale of light-front holography, $\kappa \approx 0.5$ GeV, controls the transverse dynamics of confinement in all these mesons. In heavy-heavy mesons, it also coincides numerically with the 't Hooft coupling which governs longitudinal confinement, thus reflecting the restoration of manifest $3$-dimensional rotational symmetry.

\end{abstract}
\begin{keyword}
 Light-front holography \sep 't Hooft equation \sep Hadron spectroscopy \sep Longitudinal dynamics
\end{keyword}

\end{frontmatter}

\section{Introduction} 
Although QCD is the accepted theory for the strong interactions, it is not yet possible to predict the experimentally observed hadron spectrum from first principles. This is due to our incomplete understanding of the non-perturbative aspects of QCD responsible for colour confinement. While much progress is being made with numerical simulations on the lattice \cite{Joo:2019byq}, complementary insights into non-perturbative QCD can be obtained from the AdS/CFT duality \cite{CasalderreySolana:2011us,Erdmenger:2007cm} which refers to a correspondence between gravity in a higher dimensional anti de Sitter (AdS) space and a conformal field theory (CFT) in a lower-dimensional space. The prototypical example discovered by Maldacena \cite{Maldacena:1997re} is the duality between supersymmetric $\mathrm{SU}(N_c)$ Yang-Mills theory with gauge coupling, $g_s$, in the 't Hooft limit  ($N_c \gg 1$ with $g=g_s \sqrt{N_c}$ finite) and Type IIB string theory on $\mathrm{AdS}_5 \times S_5$. The AdS/CFT correspondence may be extended to a more general gauge-gravity duality which does not simultaneously require supersymmetry, conformal symmetry and the 't Hooft limit \cite{Polchinski:2000uf,Klebanov:2000hb,Sakai:2004cn,Sakai:2005yt,Witten:1998zw,Erlich:2005qh,Brodsky:2014yha,Vega:2008te,Vega:2009zb}. 

QCD is not conformally invariant. The $\mathrm{SU}_c(3)$ invariant $(3+1)$-dim QCD Lagrangian is:
\begin{equation}
	\mathcal{L}_{\mathrm{QCD}}= \bar{\Psi} (i \gamma^\mu D_\mu -m) \Psi - \frac{1}{4} G^a_{\mu\nu}G^{a \mu \nu}
\label{QCD-L}
\end{equation}
where $D_\mu=\partial_\mu-i g_s A^{a}_\mu T^a$, and $G_{\mu \nu}^a=\partial_\mu A_\nu^a-\partial_\nu A_\mu^a + g_s c^{abc}A_\mu^b A_\nu^c$ with $[T^a, T^b]=i c^{abc} T^c$, contains two mass scales: the Higgs-generated current quark mass, $m$, and the scale $\Lambda_{\mathrm{QCD}}$ in the running coupling, generated after perturbative renormalization beyond tree-level. However, if we neglect quark masses ($m \to 0$) and ignore quantum loops (no $\Lambda_{\mathrm{QCD}}$), QCD possesses an underlying conformal symmetry. 

Light-front holography, pioneered by Brodsky and de T\'eramond \cite{Brodsky:2006uqa,deTeramond:2005su,deTeramond:2008ht,Brodsky:2014yha}, exploits this conformal limit in the Hamiltonian formulation of $(3+1)$-dim  QCD on the light front, with $N_c=3$. The valence meson light-front wavefunction then factorizes as:
\begin{equation}
	\Psi (x, \zeta, \varphi)= \frac{\phi (\zeta)}{\sqrt{2\pi \zeta}} e^{i L \varphi} X(x) 
\label{full-mesonwf}
\end{equation}
where $X(x)=\sqrt{x(1-x)}\chi(x)$ and $\zeta=\sqrt{x(1-x)} \mathbf{b}_\perp$ with $\mathbf{b}_\perp=b_\perp e^{i\varphi}$, being the transverse separation between the quark and the antiquark. $x=k^+/P^+$ is the fraction of the meson's light-front momentum, $P^+$, carried by the quark and $L$ is the orbital angular momentum quantum number. The transverse mode, $\phi(\zeta)$, satisfies the holographic Schr\"odinger Equation:
\begin{equation}
	\left(-\frac{{\rm d}^2}{{\rm d}\zeta^2}-\frac{1-4L^2}{4\zeta^2}+U_{T}(\zeta)\right)\phi(\zeta)=M_T^2 \phi(\zeta) 
	\label{Holographic-SE}
\end{equation} 
with
\begin{equation}
	U_T (\zeta) = \kappa^4 \zeta^2 + 2 \kappa^2 (J-1) 
\label{holographic-potential}
\end{equation}
where $J$ is the meson's spin. Eq. \eqref{holographic-potential} is the holographic potential at equal light-front time, $x^+=0$. While the derivation of Eq. \eqref{holographic-potential} in QCD remains an open question, its form is uniquely fixed by the underlying conformal symmetry and a holographic mapping to $\mathrm{AdS}_5$ \cite{Brodsky:2013ar}. In this mapping, the variable $\zeta$ is identified with the fifth dimension of $\mathrm{AdS}_5$, and Eq. \eqref{Holographic-SE} becomes the wave equation for spin-$J$ bosonic modes propagating in $\mathrm{AdS}_5$ spacetime distorted by a quadratic dilaton field \cite{Karch:2006pv}. The emerging hadronic scale, $\kappa$, generates the meson masses in the absence of quark masses and $\Lambda_{\mathrm{QCD}}$. The longitudinal mode, $X(x)$, is fixed by the holographic mapping of the electromagnetic (or gravitational) form factor in physical spacetime, resulting in $X(x)=\sqrt{x(1-x)}$ \cite{Brodsky:2007hb,Brodsky:2008pf}.

 The  holographic Schr\"odinger Equation admits analytical solutions: 
 \begin{equation}
 \phi_{n_T L}(\zeta) \propto \zeta^{1/2+L} \exp{\left(\frac{-\kappa^2 \zeta^2}{2}\right)}  ~ L_{n_T}^L(\kappa^2 \zeta^2) 
 \label{phi-zeta}
 \end{equation}
 and
\begin{equation}
	M_{T}^2(n_T, J, L)= 4\kappa^2 \left( n_T + \frac{J+L}{2} \right) 
\label{MT}
\end{equation}
where $J=L+S$ and $S$ is the total quark-antiquark spin, i.e. $S=0$ or $1$.\footnote{In light-front holography, the quark spin wavefunction is assumed to decouple from the confinement dynamics.} 
Importantly, Eq. \eqref{MT} predicts that the pion is massless: 
\begin{equation}
M_\pi=M_{T}(0,0,0)=0	
\label{massless-pion}
\end{equation}
just as expected in the chiral limit. Eq. \eqref{MT} also correctly predicts the Regge-like linear dependence of the meson mass squared on the radial and orbital quantum numbers.

Light-front holography needs to be extended to accommodate non-zero quark masses that generate the physical pion mass. This was originally done using a prescription by Brodsky and de T\'eramond (BdT) \cite{Brodsky:2008pg}, resulting in a first-order shift to the mass spectrum given by
\begin{align}
	\Delta M_{\mathrm{BdT}}^2 &= \int \frac{\mathrm{d} x}{x(1-x)}\nonumber\\ &\times X_{\mathrm{BdT}}^2(x) \left(\frac{m_q^2}{x}+\frac{m_{\bar{q}}^2}{1-x} \right)
\label{BT-shift}
\end{align}
where
\begin{align}
	X_{\mathrm{BdT}}(x)&= \sqrt{x(1-x)}\nonumber\\ &\times \exp\left(-\frac{(1-x)m_q^2 + xm^2_{\bar{q}}}{2\kappa^2x(1-x)} \right) 
\label{BT-X}
\end{align}
so that $M_\pi=\Delta M_{\mathrm{BdT}}$. Similarly, $M_K=\Delta M_{\mathrm{BdT}}$ when the strange quark is taken into account. Using the BdT prescription, a global fit to the spectroscopic data of light hadrons,\footnote{Supersymmetric light-front holography \cite{Brodsky:2016rvj,Brodsky:2016yod,Nielsen:2018uyn} provides a unified framework for baryons and mesons/tetraquarks.} using $m_{u/d}=0.046$ GeV and $m_s=0.357$ GeV, yields $\kappa=0.523 \pm 0.024$ GeV \cite{Brodsky:2016rvj}. The BdT prescription, together with a universal $\kappa \approx 0.5$  GeV, have been widely used in a successful phenomenology of light mesons \cite{Brodsky:2014yha,Forshaw:2012im,Ahmady:2016ujw,Ahmady:2016ufq,Ahmady:2018muv,Ahmady:2020mht,Kaur:2020emh,Gutsche:2012ez,Branz:2010ub}. The same prescription has also been used to accommodate heavy quarks, leading to the conclusion that  $\kappa \propto \sqrt{m_Q}$, where $m_Q$ is the heavy quark mass, in order to be consistent with Heavy Quark Effective Theory (HQET) \cite{PhysRevLett.66.1130} and spectroscopic data \cite{Nielsen:2018ytt,Dosch:2015bca,Dosch:2016zdv}. In other words, when the BdT prescription is used for heavy quarks, the universality of $\kappa$ seems to be lost. Refs. \cite{Gutsche:2012ez,Branz:2010ub} attempts to prevent this by using a new scale $\lambda \ne \kappa$ in Eq. \eqref{BT-X}, thus hinting at the possibility that the longitudinal mode is the solution of Schr\"odinger-like Equation different from Eq. \eqref{Holographic-SE}. 
The idea to use the 't Hooft Equation to go beyond the BdT prescription was first proposed in Ref. \cite{Chabysheva:2012fe}, with the goal of predicting the meson decay constants. Very recently, Refs. \cite{DeTeramond:2021jnn,Li:2021jqb} also go beyond the BdT prescription using a phenomenological longitudinal confinement potential, first proposed in Ref. \cite{Li:2015zda} in the context of basis light-front quantization. While both Refs. \cite{Li:2021jqb,DeTeramond:2021jnn} focus on the chiral limit and the phenomenon of chiral symmetry breaking, Ref. \cite{DeTeramond:2021jnn} extends their analysis to heavy mesons in their ground state, and discusses the relation of their approach to the 't Hooft Equation. 

In this letter, we show that the 't Hooft Equation is complementary to, and consistent with, the holographic Schr\"odinger Equation. Together, they capture the main features of $3$-dimensional confinement dynamics in (non-exotic) mesons and successfully predict their full spectrum.

\section{The '\lowercase{t} Hooft Equation}

In an earlier approach \cite{tHooft:1974pnl}, 't Hooft derived a Schr\"odinger-like equation for the longitudinal mode, starting from the QCD Lagrangian in $(1+1)$-dim in the $N_c \gg 1$ approximation. This Lagrangian now contains two mass scales: the quark mass and the gauge coupling. The resulting 't Hooft Equation is: 
\begin{align}
&\left(\frac{m_q^2}{x}+\frac{m_{\bar{q}}^2}{1-x}\right)\chi(x)\nonumber\\&\quad\quad\quad\quad\quad + U_L(x) \chi(x)=M^2_L \chi(x) \;,
  \label{tHooft}
\end{align}
with
\begin{equation}
	U_L(x)\chi(x)=\frac{g^2}{\pi} \mathcal{P} \int {\rm d}y \frac{\chi(x)-\chi(y)}{(x-y)^2}
\label{tHooft-potential}
\end{equation}
where $\mathcal{P}$ denotes the principal value prescription and $g=g_s \sqrt{N_c}$ is the (finite) 't Hooft coupling with mass dimensions which plays the role of $\Lambda_\mathrm{QCD}$. Together with the quark masses, it generates the meson masses. The 't Hooft potential, Eq. \eqref{tHooft-potential}, is derived by summing an infinite number of planar ladder and rainbow diagrams at $x^+=0$, and in the light-front gauge, $A^+=0$.   

Using the fact that, $k^+=xP^+$, is conjugate to the light-front distance, $x^-$, the Fourier transformation of Eq. \eqref{tHooft-potential} yields
\begin{equation}
	U_L(x^-)= \frac{g^2}{2} P^+ |x^-| \;,
	\label{UL-x}
\end{equation}
and, since $x^+=0$, we can rewrite Eq. \eqref{UL-x} as
\begin{equation}
	U_L(b_\parallel) = g^2 P^+ |b_\parallel| \;,
\label{UL-b}
\end{equation}
where we have chosen the notation $x^3 \equiv b_\parallel$.  Therefore, in the meson's rest frame, where $P^+=M$, the 't Hooft potential corresponds to the Coulomb potential which is linear in one space dimension.  

The end-point analysis of the 't Hooft Equation with $m_q=m_{\bar{q}}=m$, using the ansatz $\chi(x)=x^{\beta}(1-x)^{\beta}$ yields the transcendental equation:
\begin{equation}
	\frac{m^2 \pi}{g^2} - 1 + \pi \beta \cot\pi\beta=0 \;.
\label{transcendental}
\end{equation}

We note that, in the chiral limit, when $m \to 0$ with $g$ fixed, then Eq. \eqref{transcendental} implies that $\beta=0$ so that $\chi(x)=1$ and thus $X(x)=\sqrt{x(1-x)}$, i.e. the longitudinal mode of light-front holography is reproduced.
In the same limit, it is known \cite{Zhitnitsky:1985um,DeTeramond:2021jnn} that the 't Hooft Equation predicts that  $M_\pi^2 \propto m_{u/d}$, which is consistent with the  Gell-Mann-Oakes-Renner relation \cite{PhysRev.175.2195}. It is also known \cite{Grinstein:1997xk} that, in the heavy quark limit, the 't Hooft Equation predicts that $f_M \propto m_Q^{-1/2}$, as expected from Heavy-Quark-Effective-Theory (HQET) \cite{PhysRevLett.69.1018}. It is also worth noting that in a carefully constrained conformal limit, $m \to 0$ and $g \to 0$, the 't Hooft Equation possesses a gravity dual in $\mathrm{AdS}_3$ \cite{Katz:2007br}.
In our approach, these results carry over to $(3+1)$-dim since the holographic Schr\"odinger Equation gives no contribution to the pion mass (see Eq. \eqref{massless-pion}) and the meson decay constant is only sensitive to the meson wavefunction, Eq. \eqref{full-mesonwf}, evaluated at $\zeta=0$ \cite{Brodsky:2007hb}. 

Unlike the holographic Schr\"odinger Equation, the 't Hooft Equation must be solved numerically. Following Ref. \cite{Chabysheva:2012fe}, we expand the longitudinal mode onto a Jacobi polynomial basis:
\begin{equation}
\chi(x)=\sum_n c_n f_n(x)
\label{expansion-basis}	
\end{equation}
with
\begin{equation} 
f_n(x)=N_n x^{\beta_1}(1-x)^{\beta_2}P_n^{(2\beta_2,2\beta_1)}(2x-1),
 \label{eq::basis-fn}
\end{equation}
where $P_n^{(2\beta_2,2\beta_1)}$ are the Jacobi polynomials and \cite{mathematicalfunctions}
\begin{align}
	N_n=&\sqrt{(2n+\tilde{\beta}_1+\tilde{\beta}_2)}\nonumber\\
	&\times
   \sqrt{\frac{n! \Gamma(n+\tilde{\beta}_1+\tilde{\beta}_2)}{\Gamma(n+\tilde{\beta}_1+1)\Gamma(n+\tilde{\beta}_2)}}
\end{align}
with $\tilde{\beta}_1\equiv 2\beta_1$ and $\tilde{\beta}_2\equiv 2\beta_2+1$. The resulting matrix representation of Eq. \eqref{tHooft} can then be diagonalized numerically. Note that we require our predictions to be independent of the choice of basis, i.e. to remain stable with respect to variations in $\beta_{1,2}$.

\section{Predicting the meson spectrum}
\label{Predictions}

We compute the meson mass spectrum using \cite{Chabysheva:2012fe,Li:2015zda}
\begin{equation}
	M^2(n_L, n_T, J, L)= M_{T}^2(n_T, J, L) + M_L^2(n_L)
\label{tot-mass}
\end{equation}  
where $M_T^2(n_T,J,L)$ and $M_L^2(n_L)$ are the eigenvalues of  Eq. \eqref{Holographic-SE} and Eq. \eqref{tHooft} respectively. Using the light-front parity and charge conjugation operators given in Ref. \cite{Brodsky:2006ez}, we predict the parity and charge conjugation quantum numbers to the meson to be  $P=(-1)^{L+1}$ and $C=(-1)^{L+S+n_L}$ respectively. We note that $n_L \geq n_T + L$, i.e. in any hadron, an orbital and radial excitations in the transverse dynamics is always accompanied by an excitation in the longitudinal dynamics. Before showing our numerical predictions, we make two comments that are important to interpret our results.

First, we use the universal holographic mass scale: $\kappa=0.523$ GeV for all mesons. Besides the successful meson phenomenology  using similar values \cite{Forshaw:2012im,Ahmady:2016ujw,Ahmady:2016ufq,Ahmady:2018muv,Ahmady:2020mht} and the fact that it also correctly predicts $\Lambda_{\mathrm{QCD}}^{\overline{\mathrm{MS}}}$ \cite{Deur:2014qfa}, consistency with HQET also hints at a universal $\kappa$ (without fixing its numerical value). According to HQET, the masses of heavy-light pseudoscalar and vector mesons in their ground state, scale as
\begin{equation}
	M^{P/V}_{qQ} \sim m_Q 
\label{HQET-masses}
\end{equation}
with
\begin{equation}
M_{qQ}^{V} - M_{qQ}^{P} \sim \frac{1}{m_Q} \;.
\label{HQET-mass-split}
\end{equation}
Using Eq. \eqref{tot-mass}, we predict that
\begin{equation}
	M_{qQ}^V - M_{qQ}^P \sim \frac{\kappa^2}{m_Q}
\end{equation}
which is consistent with Eq. \eqref{HQET-mass-split} if $\kappa$ is universal and does not scale with $m_Q$. 

Second, we expect that $g \approx \kappa$ for heavy-heavy mesons and $g$ to deviate from $\kappa$ for light and heavy-light mesons. This can be understood starting from the general relation between a light-front and an instant-form potential in the confinement region (where kinetic energy is minimal): \cite{Trawinski:2014msa}
\begin{equation}
	U_{\mathrm{LF}}= V_{\mathrm{IF}}^2 + 4 m V_{\mathrm{IF}} \;.
	\label{LF-IF-potentials}
\end{equation}
Eq. \eqref{LF-IF-potentials} is independently true for the light-front holographic potential and the 't Hooft potential. This means that
 \begin{equation}
	U_{T} = V_\perp^2 
\label{UT-Vperp}
\end{equation}
and 
\begin{equation}
	 U_{L} = V_\parallel^2 + 4m V_\parallel \;,
\label{UL-Vpara}
\end{equation}
where $V_\perp$ and $V_\parallel$ are the instant-form potentials corresponding to the holographic and 't Hooft potentials respectively. For heavy-heavy mesons, the linear term dominates the right-hand-side of Eq. \eqref{UL-Vpara}: $U_{L} \approx 4m_Q V_\parallel$. Using Eq. \eqref{UL-b} with, $P^+=M \approx 2m_Q$, we deduce that $V_\parallel \approx (g^2/2) b_\parallel$. On the other hand, $U_T \approx (\kappa^4/4) b_\perp^2$ (since $x \approx 1/2$ in heavy-heavy mesons), and thus Eq. \eqref{UT-Vperp} implies that $V_\perp \approx (\kappa^2/2) b_\perp$. The $3$-dim rotational symmetry is restored if $V_\perp=V_\parallel$, i.e. when $g \approx \kappa$. The same arguments do not apply to heavy-light and light mesons.

 \begin{table}
 \centering 
 \begin{tabular}{  l   c   c    c  }
 \cline{1-4}
 Mesons & Light &   Heavy-light & Heavy-heavy  \\
 \cline{1-4} 
 g & 0.128 &  0.680 & 0.523  \\
 $m_{u/d}$ & 0.046 &  0.046 & - \\
 $m_s$ & 0.357 &  0.357 & -  \\
 $m_c$ & - & 1.370 & 1.370 \\
 $m_b$ & - & 4.640 & 4.640 \\
 \cline{1-4} 
\end{tabular}
\caption{The quark masses and 't Hooft couplings in $\mathrm{GeV}$. Note that we use $\kappa=0.523$ GeV for all mesons.}
\label{Tab:params}
\end{table}

\begin{figure}[hbt!]
	\begin{center}
		\includegraphics[width=\linewidth]{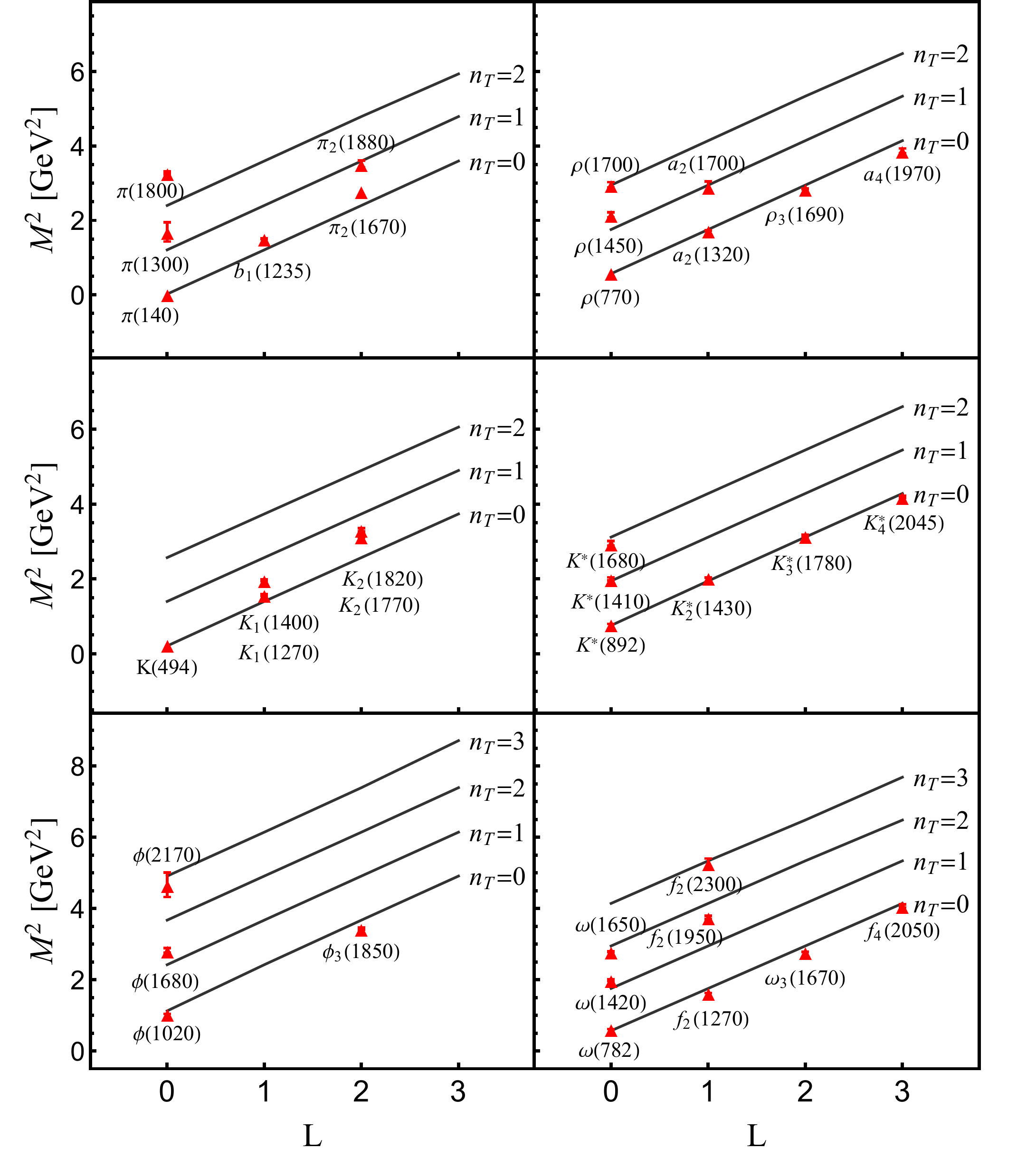}
			\caption{Our predictions for the Regge trajectories of light mesons. Data from the Particle Data Group \cite{Zyla:2020zbs}.}
		\label{Fig:light}
	\end{center}
\end{figure}

\begin{figure}[hbt!]
	\begin{center}
		\includegraphics[width=\linewidth]{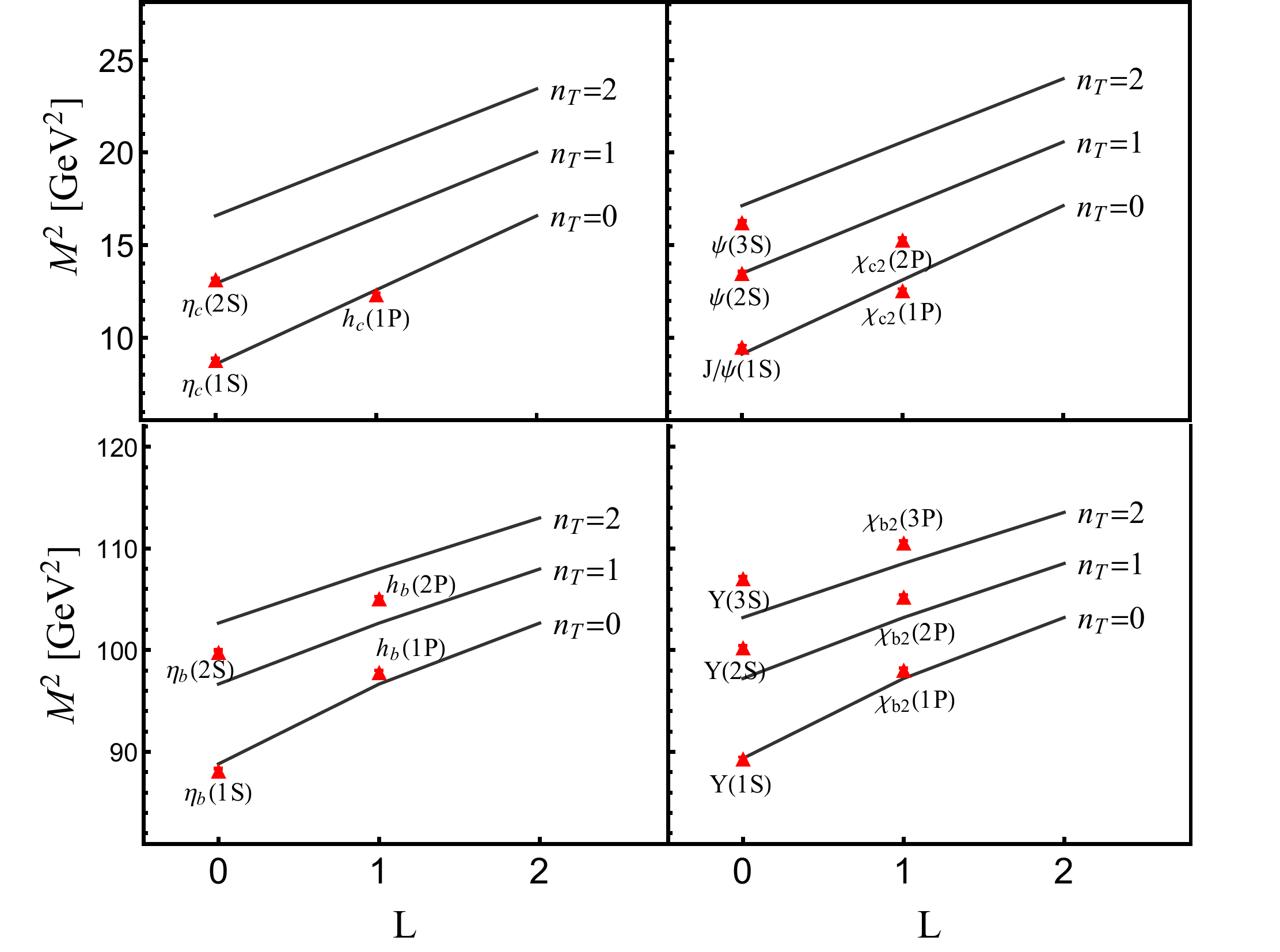}
		\caption{Our predictions for the Regge trajectories of heavy-heavy mesons. Data from the Particle Data Group \cite{Zyla:2020zbs}.}
		\label{Fig:heavy-heavy}
	\end{center}
\end{figure}

\begin{figure}[hbt!]
	\begin{center}
		\includegraphics[width=\linewidth]{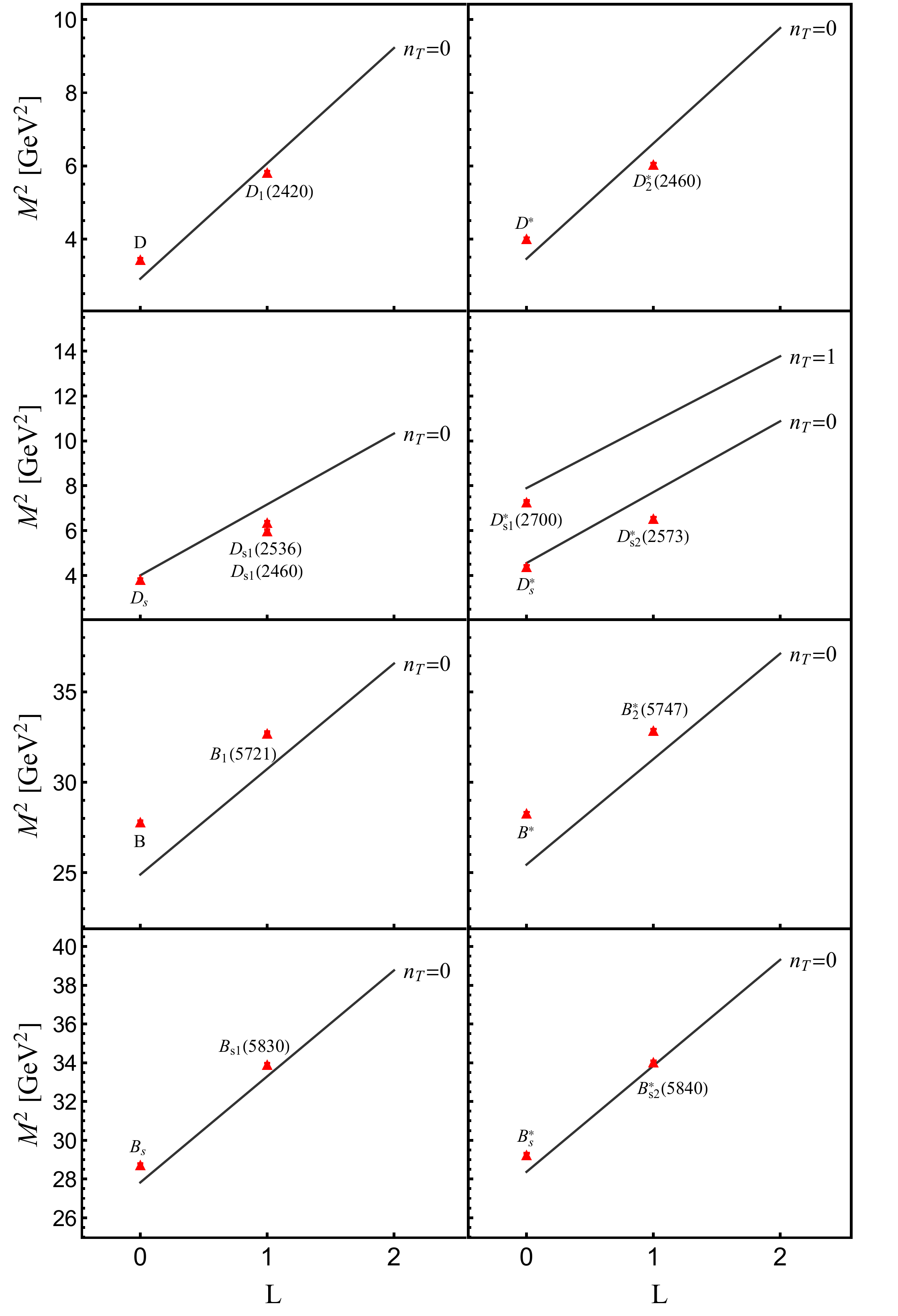}
		\caption{Our predictions for the Regge trajectories of heavy-light mesons. Data from the Particle Data Group \cite{Zyla:2020zbs}.}
		\label{Fig:heavy-light}
	\end{center}
\end{figure}

Our numerical values for quark masses and 't Hooft couplings are shown in  Table \ref{Tab:params}. Notice that we purposefully choose the same effective light quark masses as in light-front holography. The corresponding predictions for the Regge trajectories of light mesons is shown in Fig. \ref{Fig:light}. The agreement with data is very good. The quality of agreement with data is similar to that achieved using the BdT prescription for light mesons \cite{Brodsky:2014yha}. The BdT prescription can then be thought as resulting from the 't Hooft Equation with a weak longitudinal coupling, $g \ll \kappa$: see Table \ref{Tab:params}. 
Our predictions for the heavy-heavy mesons are shown in Fig. \ref{Fig:heavy-heavy}. The agreement with data is also very good, with $g=\kappa$, as we anticipated from the restoration of rotational symmetry. Finally, our predictions for heavy-light mesons are shown in Fig. \ref{Fig:heavy-light}. As can be seen, the agreement with data is good, although less impressive (even though the maximum discrepancy never exceeds $10\%$) than for the light and heavy-heavy mesons. For heavy-light mesons, the data prefer $g > \kappa$: see Table \ref{Tab:params}. Interestingly, $g$ deviates from $\kappa$ in opposite directions for light and heavy-light mesons. The underlying reason for this remains to be explored.

We emphasize that, except for pseudoscalar mesons in their ground states (for which $M_T=0$), the precise locations and slopes of the Regge trajectories of all other mesons are sensitive to both $g$ and $\kappa$. Therefore the universality of $\kappa$ across the full spectrum is non-trivial.

\section{Conclusions}
We have shown that the meson spectrum can be very well described by using the holographic Schr\"odinger Equation in conjunction with the 't Hooft Equation. We find that the emerging hadronic scale of light-front holography remains universal across the full spectrum. For heavy-heavy mesons, it coincides with the 't Hooft coupling, as expected from the restoration of manifest $3$-dimensional rotational symmetry in the non-relativistic limit.   

\section{Acknowledgements}
RS and MA are supported by individual Discovery Grants (SAPIN-2020-00051 and SAPIN-2021-00038) from the Natural Sciences and Engineering Research Council of Canada (NSERC). CM thanks the Chinese Academy of Sciences President's International Fellowship Initiative for the support via Grants No. 2021PM0023. The work of  CM is also supported by the Strategic Priority Research Program of the Chinese Academy of Sciences, Grant No. XDB34000000. The work of NS is supported by the Science and Engineering Research Board, Government of India, under the Early Career Research Award (Ref No.:ECR/2017/001905). HD thanks the Department of Science and Technology (Ref No. MTR/2019/000003) Government of India for financial support. We thank Stan Brodsky, Guy de T\'eramond and Yang Li for useful discussions.

\bibliographystyle{elsarticle-num}
\bibliography{ref.bib}

\begin{thebibliography}{10}
\expandafter\ifx\csname url\endcsname\relax
  \def\url#1{\texttt{#1}}\fi
\expandafter\ifx\csname urlprefix\endcsname\relax\def\urlprefix{URL }\fi
\expandafter\ifx\csname href\endcsname\relax
  \def\href#1#2{#2} \def\path#1{#1}\fi

\bibitem{Joo:2019byq}
B.~Jo\'o, C.~Jung, N.~H. Christ, W.~Detmold, R.~Edwards, M.~Savage,
  P.~Shanahan, {Status and Future Perspectives for Lattice Gauge Theory
  Calculations to the Exascale and Beyond}, Eur. Phys. J. A 55~(11) (2019) 199.
\newblock \href {http://arxiv.org/abs/1904.09725} {\path{arXiv:1904.09725}},
  \href {https://doi.org/10.1140/epja/i2019-12919-7}
  {\path{doi:10.1140/epja/i2019-12919-7}}.

\bibitem{CasalderreySolana:2011us}
J.~Casalderrey-Solana, H.~Liu, D.~Mateos, K.~Rajagopal, U.~A. Wiedemann,
  {Gauge/String Duality, Hot QCD and Heavy Ion Collisions}, Cambridge
  University Press, 2014.
\newblock \href {http://arxiv.org/abs/1101.0618} {\path{arXiv:1101.0618}},
  \href {https://doi.org/10.1017/CBO9781139136747}
  {\path{doi:10.1017/CBO9781139136747}}.

\bibitem{Erdmenger:2007cm}
J.~Erdmenger, N.~Evans, I.~Kirsch, E.~Threlfall, {Mesons in Gauge/Gravity Duals
  - A Review}, Eur. Phys. J. A 35 (2008) 81--133.
\newblock \href {http://arxiv.org/abs/0711.4467} {\path{arXiv:0711.4467}},
  \href {https://doi.org/10.1140/epja/i2007-10540-1}
  {\path{doi:10.1140/epja/i2007-10540-1}}.

\bibitem{Maldacena:1997re}
J.~M. Maldacena, {The Large N limit of superconformal field theories and
  supergravity}, Adv. Theor. Math. Phys. 2 (1998) 231--252.
\newblock \href {http://arxiv.org/abs/hep-th/9711200}
  {\path{arXiv:hep-th/9711200}}, \href
  {https://doi.org/10.1023/A:1026654312961}
  {\path{doi:10.1023/A:1026654312961}}.

\bibitem{Polchinski:2000uf}
J.~Polchinski, M.~J. Strassler, {The String dual of a confining
  four-dimensional gauge theory} (3 2000).
\newblock \href {http://arxiv.org/abs/hep-th/0003136}
  {\path{arXiv:hep-th/0003136}}.

\bibitem{Klebanov:2000hb}
I.~R. Klebanov, M.~J. Strassler, {Supergravity and a confining gauge theory:
  Duality cascades and chi SB resolution of naked singularities}, JHEP 08
  (2000) 052.
\newblock \href {http://arxiv.org/abs/hep-th/0007191}
  {\path{arXiv:hep-th/0007191}}, \href
  {https://doi.org/10.1088/1126-6708/2000/08/052}
  {\path{doi:10.1088/1126-6708/2000/08/052}}.

\bibitem{Sakai:2004cn}
T.~Sakai, S.~Sugimoto, {Low energy hadron physics in holographic QCD}, Prog.
  Theor. Phys. 113 (2005) 843--882.
\newblock \href {http://arxiv.org/abs/hep-th/0412141}
  {\path{arXiv:hep-th/0412141}}, \href {https://doi.org/10.1143/PTP.113.843}
  {\path{doi:10.1143/PTP.113.843}}.

\bibitem{Sakai:2005yt}
T.~Sakai, S.~Sugimoto, {More on a holographic dual of QCD}, Prog. Theor. Phys.
  114 (2005) 1083--1118.
\newblock \href {http://arxiv.org/abs/hep-th/0507073}
  {\path{arXiv:hep-th/0507073}}, \href {https://doi.org/10.1143/PTP.114.1083}
  {\path{doi:10.1143/PTP.114.1083}}.

\bibitem{Witten:1998zw}
E.~Witten, {Anti-de Sitter space, thermal phase transition, and confinement in
  gauge theories}, Adv. Theor. Math. Phys. 2 (1998) 505--532.
\newblock \href {http://arxiv.org/abs/hep-th/9803131}
  {\path{arXiv:hep-th/9803131}}, \href
  {https://doi.org/10.4310/ATMP.1998.v2.n3.a3}
  {\path{doi:10.4310/ATMP.1998.v2.n3.a3}}.

\bibitem{Erlich:2005qh}
J.~Erlich, E.~Katz, D.~T. Son, M.~A. Stephanov, {QCD and a holographic model of
  hadrons}, Phys. Rev. Lett. 95 (2005) 261602.
\newblock \href {http://arxiv.org/abs/hep-ph/0501128}
  {\path{arXiv:hep-ph/0501128}}, \href
  {https://doi.org/10.1103/PhysRevLett.95.261602}
  {\path{doi:10.1103/PhysRevLett.95.261602}}.

\bibitem{Brodsky:2014yha}
S.~J. Brodsky, G.~F. de~Teramond, H.~G. Dosch, J.~Erlich, {Light-Front
  Holographic QCD and Emerging Confinement}, Phys. Rept. 584 (2015) 1--105.
\newblock \href {http://arxiv.org/abs/1407.8131} {\path{arXiv:1407.8131}},
  \href {https://doi.org/10.1016/j.physrep.2015.05.001}
  {\path{doi:10.1016/j.physrep.2015.05.001}}.

\bibitem{Vega:2008te}
A.~Vega, I.~Schmidt, {Hadrons in AdS/QCD correspondence}, Phys. Rev. D 79
  (2009) 055003.
\newblock \href {http://arxiv.org/abs/0811.4638} {\path{arXiv:0811.4638}},
  \href {https://doi.org/10.1103/PhysRevD.79.055003}
  {\path{doi:10.1103/PhysRevD.79.055003}}.

\bibitem{Vega:2009zb}
A.~Vega, I.~Schmidt, T.~Branz, T.~Gutsche, V.~E. Lyubovitskij, {Meson wave
  function from holographic models}, Phys. Rev. D 80 (2009) 055014.
\newblock \href {http://arxiv.org/abs/0906.1220} {\path{arXiv:0906.1220}},
  \href {https://doi.org/10.1103/PhysRevD.80.055014}
  {\path{doi:10.1103/PhysRevD.80.055014}}.

\bibitem{Brodsky:2006uqa}
S.~J. Brodsky, G.~F. de~Teramond, {Hadronic spectra and light-front
  wavefunctions in holographic QCD}, Phys. Rev. Lett. 96 (2006) 201601.
\newblock \href {http://arxiv.org/abs/hep-ph/0602252}
  {\path{arXiv:hep-ph/0602252}}, \href
  {https://doi.org/10.1103/PhysRevLett.96.201601}
  {\path{doi:10.1103/PhysRevLett.96.201601}}.

\bibitem{deTeramond:2005su}
G.~F. de~Teramond, S.~J. Brodsky, {Hadronic spectrum of a holographic dual of
  QCD}, Phys. Rev. Lett. 94 (2005) 201601.
\newblock \href {http://arxiv.org/abs/hep-th/0501022}
  {\path{arXiv:hep-th/0501022}}, \href
  {https://doi.org/10.1103/PhysRevLett.94.201601}
  {\path{doi:10.1103/PhysRevLett.94.201601}}.

\bibitem{deTeramond:2008ht}
G.~F. de~Teramond, S.~J. Brodsky, {Light-Front Holography: A First
  Approximation to QCD}, Phys. Rev. Lett. 102 (2009) 081601.
\newblock \href {http://arxiv.org/abs/0809.4899} {\path{arXiv:0809.4899}},
  \href {https://doi.org/10.1103/PhysRevLett.102.081601}
  {\path{doi:10.1103/PhysRevLett.102.081601}}.

\bibitem{Brodsky:2013ar}
S.~J. Brodsky, G.~F. De~Téramond, H.~G. Dosch, {Threefold Complementary
  Approach to Holographic QCD}, Phys. Lett. B729 (2014) 3--8.
\newblock \href {http://arxiv.org/abs/1302.4105} {\path{arXiv:1302.4105}},
  \href {https://doi.org/10.1016/j.physletb.2013.12.044}
  {\path{doi:10.1016/j.physletb.2013.12.044}}.

\bibitem{Karch:2006pv}
A.~Karch, E.~Katz, D.~T. Son, M.~A. Stephanov, {Linear confinement and
  AdS/QCD}, Phys. Rev. D 74 (2006) 015005.
\newblock \href {http://arxiv.org/abs/hep-ph/0602229}
  {\path{arXiv:hep-ph/0602229}}, \href
  {https://doi.org/10.1103/PhysRevD.74.015005}
  {\path{doi:10.1103/PhysRevD.74.015005}}.

\bibitem{Brodsky:2007hb}
S.~J. Brodsky, G.~F. de~Teramond, {Light-Front Dynamics and AdS/QCD
  Correspondence: The Pion Form Factor in the Space- and Time-Like Regions},
  Phys. Rev. D 77 (2008) 056007.
\newblock \href {http://arxiv.org/abs/0707.3859} {\path{arXiv:0707.3859}},
  \href {https://doi.org/10.1103/PhysRevD.77.056007}
  {\path{doi:10.1103/PhysRevD.77.056007}}.

\bibitem{Brodsky:2008pf}
S.~J. Brodsky, G.~F. de~Teramond, {Light-Front Dynamics and AdS/QCD
  Correspondence: Gravitational Form Factors of Composite Hadrons}, Phys. Rev.
  D78 (2008) 025032.
\newblock \href {http://arxiv.org/abs/0804.0452} {\path{arXiv:0804.0452}},
  \href {https://doi.org/10.1103/PhysRevD.78.025032}
  {\path{doi:10.1103/PhysRevD.78.025032}}.

\bibitem{Brodsky:2008pg}
S.~J. Brodsky, G.~F. de~Teramond, {AdS/CFT and Light-Front QCD}, Subnucl. Ser.
  45 (2009) 139--183.
\newblock \href {http://arxiv.org/abs/0802.0514} {\path{arXiv:0802.0514}},
  \href {https://doi.org/10.1142/9789814293242_0008}
  {\path{doi:10.1142/9789814293242_0008}}.

\bibitem{Brodsky:2016rvj}
S.~J. Brodsky, G.~F. de~Teramond, H.~G. Dosch, C.~Lorce,
  {Meson/Baryon/Tetraquark Supersymmetry from Superconformal Algebra and
  Light-Front Holography}, Int. J. Mod. Phys. A31~(19) (2016) 1630029.
\newblock \href {http://arxiv.org/abs/1606.04638} {\path{arXiv:1606.04638}},
  \href {https://doi.org/10.1142/S0217751X16300295}
  {\path{doi:10.1142/S0217751X16300295}}.

\bibitem{Brodsky:2016yod}
S.~J. Brodsky, G.~F. de~T\'eramond, H.~G. Dosch, C.~Lorc\'e, {Universal
  Effective Hadron Dynamics from Superconformal Algebra}, Phys. Lett. B 759
  (2016) 171--177.
\newblock \href {http://arxiv.org/abs/1604.06746} {\path{arXiv:1604.06746}},
  \href {https://doi.org/10.1016/j.physletb.2016.05.068}
  {\path{doi:10.1016/j.physletb.2016.05.068}}.

\bibitem{Nielsen:2018uyn}
M.~Nielsen, S.~J. Brodsky, {Hadronic superpartners from a superconformal and
  supersymmetric algebra}, Phys. Rev. D 97~(11) (2018) 114001.
\newblock \href {http://arxiv.org/abs/1802.09652} {\path{arXiv:1802.09652}},
  \href {https://doi.org/10.1103/PhysRevD.97.114001}
  {\path{doi:10.1103/PhysRevD.97.114001}}.

\bibitem{Forshaw:2012im}
J.~R. Forshaw, R.~Sandapen, {An AdS/QCD holographic wavefunction for the rho
  meson and diffractive rho meson electroproduction}, Phys. Rev. Lett. 109
  (2012) 081601.
\newblock \href {http://arxiv.org/abs/1203.6088} {\path{arXiv:1203.6088}},
  \href {https://doi.org/10.1103/PhysRevLett.109.081601}
  {\path{doi:10.1103/PhysRevLett.109.081601}}.

\bibitem{Ahmady:2016ujw}
M.~Ahmady, R.~Sandapen, N.~Sharma, {Diffractive $\rho$ and $\phi$ production at
  HERA using a holographic AdS/QCD light-front meson wave function}, Phys. Rev.
  D94~(7) (2016) 074018.
\newblock \href {http://arxiv.org/abs/1605.07665} {\path{arXiv:1605.07665}},
  \href {https://doi.org/10.1103/PhysRevD.94.074018}
  {\path{doi:10.1103/PhysRevD.94.074018}}.

\bibitem{Ahmady:2016ufq}
M.~Ahmady, F.~Chishtie, R.~Sandapen, {Spin effects in the pion holographic
  light-front wavefunction}, Phys. Rev. D 95~(7) (2017) 074008.
\newblock \href {http://arxiv.org/abs/1609.07024} {\path{arXiv:1609.07024}},
  \href {https://doi.org/10.1103/PhysRevD.95.074008}
  {\path{doi:10.1103/PhysRevD.95.074008}}.

\bibitem{Ahmady:2018muv}
M.~Ahmady, C.~Mondal, R.~Sandapen, {Dynamical spin effects in the holographic
  light-front wavefunctions of light pseudoscalar mesons}, Phys. Rev. D98~(3)
  (2018) 034010.
\newblock \href {http://arxiv.org/abs/1805.08911} {\path{arXiv:1805.08911}},
  \href {https://doi.org/10.1103/PhysRevD.98.034010}
  {\path{doi:10.1103/PhysRevD.98.034010}}.

\bibitem{Ahmady:2020mht}
M.~Ahmady, S.~Kaur, C.~Mondal, R.~Sandapen, {Light-front holographic radiative
  transition form factors for light mesons}, Phys. Rev. D102~(3) (2020) 034021.
\newblock \href {http://arxiv.org/abs/2006.07675} {\path{arXiv:2006.07675}},
  \href {https://doi.org/10.1103/PhysRevD.102.034021}
  {\path{doi:10.1103/PhysRevD.102.034021}}.

\bibitem{Kaur:2020emh}
S.~Kaur, C.~Mondal, H.~Dahiya, {Light-front holographic $\rho$-meson
  distributions in the momentum space}, JHEP 01 (2021) 136.
\newblock \href {http://arxiv.org/abs/2009.04288} {\path{arXiv:2009.04288}},
  \href {https://doi.org/10.1007/JHEP01(2021)136}
  {\path{doi:10.1007/JHEP01(2021)136}}.

\bibitem{Gutsche:2012ez}
T.~Gutsche, V.~E. Lyubovitskij, I.~Schmidt, A.~Vega, {Chiral Symmetry Breaking
  and Meson Wave Functions in Soft-Wall AdS/QCD}, Phys. Rev. D 87~(5) (2013)
  056001.
\newblock \href {http://arxiv.org/abs/1212.5196} {\path{arXiv:1212.5196}},
  \href {https://doi.org/10.1103/PhysRevD.87.056001}
  {\path{doi:10.1103/PhysRevD.87.056001}}.

\bibitem{Branz:2010ub}
T.~Branz, T.~Gutsche, V.~E. Lyubovitskij, I.~Schmidt, A.~Vega, {Light and heavy
  mesons in a soft-wall holographic approach}, Phys. Rev. D82 (2010) 074022.
\newblock \href {http://arxiv.org/abs/1008.0268} {\path{arXiv:1008.0268}},
  \href {https://doi.org/10.1103/PhysRevD.82.074022}
  {\path{doi:10.1103/PhysRevD.82.074022}}.

\bibitem{PhysRevLett.66.1130}
N.~Isgur, M.~B. Wise,
  \href{https://link.aps.org/doi/10.1103/PhysRevLett.66.1130}{Spectroscopy with
  heavy-quark symmetry}, Phys. Rev. Lett. 66 (1991) 1130--1133.
\newblock \href {https://doi.org/10.1103/PhysRevLett.66.1130}
  {\path{doi:10.1103/PhysRevLett.66.1130}}.
\newline\urlprefix\url{https://link.aps.org/doi/10.1103/PhysRevLett.66.1130}

\bibitem{Nielsen:2018ytt}
M.~Nielsen, S.~J. Brodsky, G.~F. de~T\'eramond, H.~G. Dosch, F.~S. Navarra,
  L.~Zou, {Supersymmetry in the Double-Heavy Hadronic Spectrum}, Phys. Rev. D
  98~(3) (2018) 034002.
\newblock \href {http://arxiv.org/abs/1805.11567} {\path{arXiv:1805.11567}},
  \href {https://doi.org/10.1103/PhysRevD.98.034002}
  {\path{doi:10.1103/PhysRevD.98.034002}}.

\bibitem{Dosch:2015bca}
H.~G. Dosch, G.~F. de~Teramond, S.~J. Brodsky, {Supersymmetry Across the Light
  and Heavy-Light Hadronic Spectrum}, Phys. Rev. D 92~(7) (2015) 074010.
\newblock \href {http://arxiv.org/abs/1504.05112} {\path{arXiv:1504.05112}},
  \href {https://doi.org/10.1103/PhysRevD.92.074010}
  {\path{doi:10.1103/PhysRevD.92.074010}}.

\bibitem{Dosch:2016zdv}
H.~G. Dosch, G.~F. de~Teramond, S.~J. Brodsky, {Supersymmetry Across the Light
  and Heavy-Light Hadronic Spectrum II}, Phys. Rev. D 95~(3) (2017) 034016.
\newblock \href {http://arxiv.org/abs/1612.02370} {\path{arXiv:1612.02370}},
  \href {https://doi.org/10.1103/PhysRevD.95.034016}
  {\path{doi:10.1103/PhysRevD.95.034016}}.

\bibitem{Chabysheva:2012fe}
S.~S. Chabysheva, J.~R. Hiller, {Dynamical model for longitudinal wave
  functions in light-front holographic QCD}, Annals Phys. 337 (2013) 143--152.
\newblock \href {http://arxiv.org/abs/1207.7128} {\path{arXiv:1207.7128}},
  \href {https://doi.org/10.1016/j.aop.2013.06.016}
  {\path{doi:10.1016/j.aop.2013.06.016}}.

\bibitem{DeTeramond:2021jnn}
G.~F. De~T\'eramond, S.~J. Brodsky, {Longitudinal dynamics and chiral symmetry
  breaking in holographic light-front QCD} (3 2021).
\newblock \href {http://arxiv.org/abs/2103.10950} {\path{arXiv:2103.10950}}.

\bibitem{Li:2021jqb}
Y.~Li, J.~P. Vary, {Light-front holography with chiral symmetry breaking} (3
  2021).
\newblock \href {http://arxiv.org/abs/2103.09993} {\path{arXiv:2103.09993}}.

\bibitem{Li:2015zda}
Y.~Li, P.~Maris, X.~Zhao, J.~P. Vary, {Heavy Quarkonium in a Holographic
  Basis}, Phys. Lett. B 758 (2016) 118--124.
\newblock \href {http://arxiv.org/abs/1509.07212} {\path{arXiv:1509.07212}},
  \href {https://doi.org/10.1016/j.physletb.2016.04.065}
  {\path{doi:10.1016/j.physletb.2016.04.065}}.

\bibitem{tHooft:1974pnl}
G.~'t~Hooft, {A Two-Dimensional Model for Mesons}, Nucl. Phys. B 75 (1974)
  461--470.
\newblock \href {https://doi.org/10.1016/0550-3213(74)90088-1}
  {\path{doi:10.1016/0550-3213(74)90088-1}}.

\bibitem{Zhitnitsky:1985um}
A.~R. Zhitnitsky, {On Chiral Symmetry Breaking in {QCD} in Two-dimensions ($N_c
  \to$ Infinity)}, Phys. Lett. B 165 (1985) 405--409.
\newblock \href {https://doi.org/10.1016/0370-2693(85)91255-9}
  {\path{doi:10.1016/0370-2693(85)91255-9}}.

\bibitem{PhysRev.175.2195}
M.~Gell-Mann, R.~J. Oakes, B.~Renner,
  \href{https://link.aps.org/doi/10.1103/PhysRev.175.2195}{Behavior of current
  divergences under
  ${\mathrm{su}}_{3}\ifmmode\times\else\texttimes\fi{}{\mathrm{su}}_{3}$},
  Phys. Rev. 175 (1968) 2195--2199.
\newblock \href {https://doi.org/10.1103/PhysRev.175.2195}
  {\path{doi:10.1103/PhysRev.175.2195}}.
\newline\urlprefix\url{https://link.aps.org/doi/10.1103/PhysRev.175.2195}

\bibitem{Grinstein:1997xk}
B.~Grinstein, R.~F. Lebed, {Explicit quark - hadron duality in heavy - light
  meson weak decays in the 't Hooft model}, Phys. Rev. D 57 (1998) 1366--1378.
\newblock \href {http://arxiv.org/abs/hep-ph/9708396}
  {\path{arXiv:hep-ph/9708396}}, \href
  {https://doi.org/10.1103/PhysRevD.57.1366}
  {\path{doi:10.1103/PhysRevD.57.1366}}.

\bibitem{PhysRevLett.69.1018}
B.~Grinstein, P.~F. Mende,
  \href{https://link.aps.org/doi/10.1103/PhysRevLett.69.1018}{Heavy mesons in
  two dimensions}, Phys. Rev. Lett. 69 (1992) 1018--1021.
\newblock \href {https://doi.org/10.1103/PhysRevLett.69.1018}
  {\path{doi:10.1103/PhysRevLett.69.1018}}.
\newline\urlprefix\url{https://link.aps.org/doi/10.1103/PhysRevLett.69.1018}

\bibitem{Katz:2007br}
E.~Katz, T.~Okui, {The 't Hooft model as a hologram}, JHEP 01 (2009) 013.
\newblock \href {http://arxiv.org/abs/0710.3402} {\path{arXiv:0710.3402}},
  \href {https://doi.org/10.1088/1126-6708/2009/01/013}
  {\path{doi:10.1088/1126-6708/2009/01/013}}.

\bibitem{mathematicalfunctions}
M.~Abramowitz, I.~A. Stegun, Handbook of Mathematical Functions, 1965.

\bibitem{Brodsky:2006ez}
S.~J. Brodsky, S.~Gardner, D.~S. Hwang, {Discrete symmetries on the light front
  and a general relation connecting nucleon electric dipole and anomalous
  magnetic moments}, Phys. Rev. D 73 (2006) 036007.
\newblock \href {http://arxiv.org/abs/hep-ph/0601037}
  {\path{arXiv:hep-ph/0601037}}, \href
  {https://doi.org/10.1103/PhysRevD.73.036007}
  {\path{doi:10.1103/PhysRevD.73.036007}}.

\bibitem{Deur:2014qfa}
A.~Deur, S.~J. Brodsky, G.~F. de~Teramond, {Connecting the Hadron Mass Scale to
  the Fundamental Mass Scale of Quantum Chromodynamics}, Phys. Lett. B 750
  (2015) 528--532.
\newblock \href {http://arxiv.org/abs/1409.5488} {\path{arXiv:1409.5488}},
  \href {https://doi.org/10.1016/j.physletb.2015.09.063}
  {\path{doi:10.1016/j.physletb.2015.09.063}}.

\bibitem{Trawinski:2014msa}
A.~P. Trawi\'nski, S.~D. G\l{}azek, S.~J. Brodsky, G.~F. de~T\'eramond, H.~G.
  Dosch, {Effective confining potentials for QCD}, Phys. Rev. D 90~(7) (2014)
  074017.
\newblock \href {http://arxiv.org/abs/1403.5651} {\path{arXiv:1403.5651}},
  \href {https://doi.org/10.1103/PhysRevD.90.074017}
  {\path{doi:10.1103/PhysRevD.90.074017}}.

\bibitem{Zyla:2020zbs}
P.~Zyla, et~al., {Review of Particle Physics}, PTEP 2020~(8) (2020) 083C01.
\newblock \href {https://doi.org/10.1093/ptep/ptaa104}
  {\path{doi:10.1093/ptep/ptaa104}}.

\end{thebibliography}
\end{document}